\documentclass[epj]{svjour}
\usepackage{graphics}
\usepackage{amssymb}
\usepackage{amsmath}
\begin{document}
\title{Plane symmetric model in $f(R,T)$ gravity}
\author{Vijay Singh\inst{1} \and Aroonkumar Beesham\inst{2,1}
}                     
\offprints{Vijay Singh}   
\institute{Department of Mathematical Sciences,
 University of Zululand,
 KwaDlangezwa - 3886,
 South Africa\\
\email{gtrcosmo@gmail.com}
\and Faculty of Natural Sciences,
Mangosuthu University of Technology,
Umlazi,
South Africa\\
\email{abeesham@yahoo.com}
}
\date{Received: date / Revised version: date}
%
\abstract{
A plane symmetric Bianchi-I model is explored in $f(R,T)$ gravity, where $R$ is the Ricci scalar and $T$ is the trace of energy-momentum tensor. The solutions are obtained with the consideration of a specific Hubble parameter which yields a constant deceleration parameter. The various evolutionary phases are identified under the constraints obtained for physically viable cosmological scenarios. Although a single (primary) matter source is taken, due to the coupling between matter and $f(R,T)$ gravity, an additional matter source appears, which mimics a perfect fluid or exotic matter. The solutions are also extended to the case of a scalar field model. The kinematical behavior of the model remains independent of $f(R,T)$ gravity. The physical behavior of the effective matter also remains the same as in general relativity. It is found that $f(R,T)$ gravity can be a good alternative to the hypothetical candidates of dark energy to describe the present accelerating expansion of the universe.
} 
\maketitle
\section{Introduction}
\label{intro}
The expansion of the universe is in an accelerating phase at present. Many attempts have been posed to explain this unexpected phenomenon but none of them is compelling. All these attempts are broadly divided into two categories: dark energy (DE), and modified theories of gravity. A number of candidates have been proposed for DE, many existing modified gravity theories have been employed and some new modifications have been proposed to explain this mysterious behavior of the universe. The progress in developing new theories is still going on. The basic idea of DE is the hypothesis of some exotic matter featuring an extraordinary characteristic of anti-gravity, which due to highly negative pressure, generates a repulsive force to speed up the expansion of the universe (for details see the review \cite{Bambaetal2012}, and references therein). On the other hand, modified theories of gravity entail a modification of Einstein's general relativity (GR) in some or other way (for details see the review \cite{NojiriOdintsov2011}, and references therein). In most modified theories, the geometry and matter Lagrangian have an additive structure. However, there is no fundamental principle considering so. Therefore, working in the direction of development of a general non-minimal coupling between matter and geometry, Harko {\it et al.} \cite{Harkoetal2011}, in 2011, introduced $f(R,T)$ gravity, where $f(R,T)$ is an arbitrary function of the Ricci scalar $R$, and the trace $T$ of the energy-momentum tensor. A prominent feature of this theory is that an extra acceleration is always present that results not only from a geometrical contribution, but also from the matter content. This interesting phenomena provides some significant signatures and effects which could distinguish and discriminate between this theory and other  gravitational theories. Motivating by this fact many researchers have  reconstructed, tested and implemented $f(R,T)$ gravity on galactic and intra-galactic scales (see for example \cite{Jamiletal2012,HoundjoPiattella2012,Alvarengaetal2013jmp4,Azizi2013,Alvarengaetal2013prd87,Sharifetal2013epjp128,Chakraborty2013,Houndjoetal2013cjp91,ShabaniFarhoudiPRD2013,Pasquaetal2013,KumarSinghASS2015,Baffouetal2015,SantosFerstMPLA2015,NoureenZubairEPJC2015,ZubairNoureenEPJC2015,NoureenetalEPJC2015,SinghSingh2016,AlhamzawiAlhamzawiIJMPD2016,SalehiAftabiJHEP2016,MomenietalASS2016,Alvesetal2016,YousafetalPRD2016,MoraesetalEPJC2018,ZubairetalEPJC2016,DasetalEPJC2016,SahooetalEPJC2018,SinghBeeshamEPJC2018,SrivastavaSinghASS2018,SharifAnwarASS2018,ShabaniZiaieEPJC2018,Bhattietal1709.06892,RajabiNozariPRD2017,MoraesetalIJMPD2019,LobatoetalEPJP2019,BaffouetalPRD2018,DebetalPRD2018,DebetalMNRAS2019,TretyakovEPJC2019,ElizaldeKhurshudyanPRD2018,OrdinesCarlsonPRD2019,MauryaTello-OrtizbJCAP2019,DebnathIJGMMP2019,Mauryaetal1907.05209,SahooBhattacharjee1907.13460} and references therein). Most of these works have been done in an isotropic background. Though our universe is approximated as homogenous and isotropic at large scales, there are theoretical arguments for the existence of an anisotropic phase at early stages of evolution that approaches an isotropic one at late times. The outcomes of various observational data also support this anticipation. Also, it is highly unlikely that the universe is exactly isotropic.

While the first detection of the cosmic microwave background (CMB) radiation revealed a homogeneous and isotropic background glow across the sky, the improvements in instrumentation and increasingly precise measurements of the background radiation led to the discovery that the CMB radiation was actually slightly anisotropic. The observational data of the CMB \cite{Netterfieldetal2002} and Wilkinson Microwave Anisotropy Probe (WMAP) \cite{BennettetalApJ2013,Hinshawetal2013} reveal that the universe on small scales is somewhat inhomogeneous and anisotropic. The recent outcomes of the Planck Collaboration \cite{PlanckCollaboration2018} also evidence tiny anisotropies in temperature. The small-scale anisotropies in the CMB radiation are dominated by the acoustic oscillations created during recombination. Therefore, it is supposed that the initially small-amplitude CMB anisotropies at recombination seeded the large-scale discrete structures, e.g., galaxy clusters, filaments, and voids that we see today. As these anisotropies grow over time under the influence of gravity, the dynamics of their evolution becomes increasingly more complicated. While the amplitudes are small (linear), the evolution can be predicted analytically. Consequently, the anisotropic models play a significant role to describe the behavior of the early universe. Therefore, the study of early anisotropic phases of the universe gains a lot of interest, and $f(R,T)$ gravity is no exception. Many authors have considered homogenous cosmological models in anisotropic and higher dimensional space-times as well (see for example  \cite{SharifZubair2013arxiv,ShamirJETP2014,ShamirEPJC2015,TiwariBeeshamASS2018,EsmaeiliJHEP2018,ReddyetalIJTP2012,SharifZubair2013jpsj,RamPriyankaASS2013,MoraesEPJC2015,SofuogluASS2016} and references therein).

Amongst the various families of homogeneous but anisotropic cosmological models, the study of the possible effects of anisotropy at early times makes the Bianchi type-I (B-I) model as the prime alternative as it is the simplest. The first work of $f(R,T)$ gravity in anisotropic spacetime was considered by Adhav \cite{AdhavASS2012}. Considering a particular form $f(R,T)=R+2\lambda T$, where $\lambda$ is an arbitrary constant, and assuming a constant expansion rate, the author obtained solutions in a locally-rotationally-symmetric (LRS) B-I spacetime model. However, due to an incorrect field equation, the solutions presented by him are mathematically and hence physically invalid. We have reconsidered Adhav's work elsewhere \cite{SinghBeeshamApSS2019} where we have presented the correct field equations and their solutions, and also explored the geometrical and physical properties thoroughly. Later on, a number of authors \cite{SharifZubair2013arxiv,ShamirJETP2014,ShamirEPJC2015,TiwariBeeshamASS2018,EsmaeiliJHEP2018} have considered cosmological models in B-I space-times.

Following Adhav's work, Sahoo {\it et al.} \cite{SahooetalEPJP2014} considered an axially symmetric B-I model with a specific law of the Hubble parameter. Although the authors started with an axially symmetric B-I metric, due to an assumption to simplify the field equations, the final field equations appearing in their work become identical to those of a plane symmetric B-I model. Notwithstanding, even if it is supposed that their work was done in a plane symmetric B-I spacetime, another serious issue in their formulation is the wrong signs in the field equations. Their solutions are also therefore mathematically incorrect.

Our purpose in the present work is to address the correct field equations of the model considered by Sahoo {\it et al.} \cite{SahooetalEPJP2014} and to explore the geometrical and physical behaviors thoroughly. Since the field equations solved by Sahoo {\it et al.} \cite{SahooetalEPJP2014} are identical to the plane symmetric B-I model,  rather than starting with an axially symmetric B-I model, we directly consider a plane symmetric B-I model. Although we consider a single matter content in our model, due to the coupling terms of trace $T$ with the matter an extra matter appears in the field equations. We name this extra matter as coupled matter. First, we obtain the constraints for the primary matter to obey the weak energy condition (WEC). This not only ensures a realistic cosmological scenario, but also helps to identify the various evolutionary phases of the universe, specifically, it distinguishes between early inflation and late time acceleration. Anticipating the dual nature of primary matter, we replace it with a scalar field to know the actual nature (perfect fluid, quintessence or phantom) of primary matter in the various evolutionary phases. Further, we analyze the behavior of coupled matter in the various evolutionary phases. In this way,  we determine which matter source causes inflation, deceleration and late time acceleration, and what is the role of $f(R,T)$ gravity in this model.

The work is organised as follows. In Sect. 2, we show that the geometrical behavior of the model reported by Sahoo {\it et al.} \cite{SahooetalEPJP2014} is independent of $f(R,T)$ gravity. In Sect. 3 we present the correct field equations for a plane symmetric B-I spacetime in $f(R,T)=R+2f(T)$ gravity followed by the field equations for $f(T)=\lambda T$. In Sect. 3.1, we study the model for $m\neq3$. The constraints are found for a physically realistic scenario, and the possibility of various evolutionary phases which may occur under the obtained constraints is investigated in the  different cases. The scalar field model is considered thereafter in Sect. 3.2. The nature of coupled matter is examined in Sect. 3.3. Sect. 3.4 is devoted to the particular case when $m=3$. The findings are accumulated in the concluding Sect. 4. Note that wheresoever cited, the equation numbers and the section numbers within inverted commas refer to the Ref. \cite{SahooetalEPJP2014}, and otherwise, they refer to our work.

\section{The model in Einstein's gravity}
\label{sec:1}
The spatially homogenous and anisotropic plane symmetric B-I spacetime metric is given by
\begin{equation}
ds^{2} =dt^{2}-A^2(dx^2+dy^2)-B^2dz^2,
\end{equation}
where $A$ and $B$ are the scale factors, and are functions of cosmic time $t$.
The average scale factor for the metric (1) is defined as
\begin{equation}
  a=(A^2B)^{\frac{1}{3}}.
\end{equation}
The average Hubble parameter (average expansion rate) $H$, which is the generalization of the Hubble parameter in the isotropic case, is given by
\begin{equation}
  H=\frac{1}{3}\left(2\frac{\dot A}{A}+\frac{\dot B}{B}\right),
\end{equation}
where a dot denotes the derivative with respect to $t$.

The energy-momentum tensor of the matter is given as
\begin{equation}
  T_{\mu\nu}=(\rho+p)u_\mu u_\nu-pg_{\mu\nu},
\end{equation}
\noindent where $\rho$ is the energy density and $p$ is the thermodynamical pressure of the effective matter. In comoving coordinates $u^\mu=\delta_0^\mu$, where $u_\mu$ is the four-velocity of the fluid which satisfies the condition $u_\mu u^\nu=1$.

The Einstein field equations read as
\begin{equation}
R_{\mu\nu}-\frac{1}{2}R g_{\mu\nu}=T_{\mu\nu},
\end{equation}
where the system of units $8\pi G=1=c$ is used.

The above field equations for the metric (1) and energy-momentum tensor (4) yield
\begin{eqnarray}
  \left(\frac{\dot A}{A}\right)^2+2\frac{\dot A\dot B}{A B}&=&\rho,\\
2\frac{\ddot A}{ A}+\left(\frac{\dot A}{A}\right)^2&=&-p,\\
\frac{\ddot A}{A}+\frac{\ddot B}{B}+\frac{\dot A \dot B}{AB}&=&-p.
\end{eqnarray}
These are three independent equations with four unknowns namely $A$, $B$, $\rho$ and $p$. Therefore, one requires a supplementary constraint to find exact solutions of the field equations. Sahoo {\it et al.} \cite{SahooetalEPJP2014} considered
\begin{equation}
H=k(A^2B)^{-\frac{m}{3}},
\end{equation}
where $k>0$ and $ m\geq0$ are constants. The authors worked on two cases, i.e., $m=0$ and $m\neq0$. It is to be noted that Adhav \cite{AdhavASS2012} considered an LRS B-I model in $f(R,T)$ gravity with a similar assumption for the case $m=0$. However, the solutions presented by Adhav are not mathematically valid due to the consideration of a wrong field equation. We have addressed this issue elsewhere \cite{SinghBeeshamApSS2019}. In Cartesian coordinates, the line element of a plane-symmetric B-I spacetime is indistinguishable from that of an LRS B-I spacetime. Consequently, if one solves the field equations of both models under a  common assumption, only the geometrical behavior is reverted directional wise, but the physical behavior remains the same \cite{BeeshamSingh2002.08654}. So we shall not consider the case $m=0$ in the present study for the sake of avoiding repetition. However, it would be worthwhile to discuss an important point here that although the geometrical parameters worked out by Sahoo {\it et al.} \cite{SahooetalEPJP2014} are mathematically correct, we note in section ``4.2'' that they emphasized the behavior of the model at $t=0$, while the universe has an infinite past in this case. The authors probably misunderstood the origin of the universe. They considered the origin of the universe at $t=0$. More precisely, all the geometrical and physical quantities in case of $m=0$ are finite at any instance of time, but the authors mentioned that these quantities are finite only at $t=0$. The readers may refer our recent works \cite{SinghBeeshamApSS2019,SinghBeeshamGRG2019} to gain more details on this issue.

Let us now proceed with $m\neq0$. The deceleration parameter, $q=-a\ddot a/\dot a^2=-1-\dot H/ H^2$, takes a constant value
\begin{equation}
q=m-1.
\end{equation}
The models with $m<1$ correspond to accelerated universes whereas the models with $m>1$ correspond to decelerated universes.

From (7) and (8), one has
\begin{equation}
\frac{\dot A}{A} -\frac{\dot B}{B}=\frac{\beta}{A^2B},
\end{equation}
where $\beta$ is a constant of integration.

From (3) and (11), by the use of (9), one obtains
\begin{eqnarray}
  A&=&\left\{
  \begin{array}{ll}
     c_1 t^\frac{1}{m} \exp{\left[{\frac{\beta  m t (k m t)^{-\frac{3}{m}}}{3 (m-3)}}\right]}; & \hbox{$m\neq3$,} \\
    c_1 t^{\frac{\beta +3 k}{9 k}}; & \hbox{$m=3$,}
  \end{array}
\right.\\
B&=&\left\{
  \begin{array}{ll}
     c_1t^\frac{1}{m} \exp{\left[{-{\frac{2 \beta  m t (k m t)^{-\frac{3}{m}}}{3(m-3)}}}\right]}; & \hbox{$m\neq3$,} \\
    c_1 t^{\frac{3 k-2 \beta }{9 k}}; & \hbox{$m=3$,}
  \end{array}
\right.
\end{eqnarray}
where $c_1$ is an integration constant. It is to be noted that Sahoo {\it et al.} \cite{SahooetalEPJP2014} neglected the solutions for $m=3$.

In section ``4.4", namely, ``Physical behavior of the model", Sahoo {\it et al.} \cite{SahooetalEPJP2014} discussed merely some kinematical properties. Since all the kinematical parameters are obtained by the use of scale factors (12) and (13) which are independent of $f(R,T)$ gravity, therefore, the geometrical behavior of the model remains identical to one in GR \cite{SinghBeeshamGRG2019}. Similarly, in section ``4.4" on the basis of energy density and pressure diverging to infinity at $t=0$, they mentioned that it indicates a big-bang singularity. However, this is also an outcome of the assumption (9). It is to be noted that all the solutions under the assumption (9) correspond to non-singular models when $m=0$ and singular models when $m\neq0$ (see \cite{SinghBeeshamGRG2019} and references therein).

What is more preposterous, in the concluding section ``5", Sahoo {\it et al.} \cite{SahooetalEPJP2014} repeated some general features of $f(R,T)$ gravity theory which are not the outcomes of their study. The authors added, these models represent the accelerated expansion of the universe. However, the models with $m>1$ correspond to decelerated universes. In fact, an acceleration or a deceleration is guaranteed from the Hubble parameter assumed to obtain the solutions. The Hubble parameter (9) leads to the constant deceleration parameter (10). Consequently, one may fix the expansion of the universe to be decelerated or accelerated just by choosing an appropriate value of $m$ as per desired evolution. Indeed, the acceleration or deceleration in such formulations becomes independent of the theory. Thus, we see that although Sahoo {\it et al.} \cite{SahooetalEPJP2014} considered $f(R,T)$ gravity, but they have not discussed the role of modified gravity in their work. All the results presented by them remain the same as in GR.

Let us explain the way of studying the models where the geometrical behavior of the universe is fixed by the scale factor or by any other geometrical parameters. The main objective of such studies must be:
\begin{description}
  \item[(i)] To investigate the cause which can give the desired geometrical behavior.
  \item[(ii)] To identify the matter in the presence of which the model can depict the desired evolution.
  \item[(iii)] To examine the role of the modified gravity theory (if working in any such theory).
  \item[(iv)] To differentiate the outcomes from that in GR.
\end{description}
Before all of these, the most important thing is to ensure whether the adhoc assumption made to obtain the solutions is consistent to give a viable cosmological scenario. In other words, the energy density of matter must be positive. This procedure has been followed in many earlier works \cite{BenerjeeDasGRG2005,AkarsuDereliIJTP2012,SinghSinghIJTP2012,SinghSinghASS2012,SinghSinghASS2013,SinghSinghGRG2014,SinghSinghASS2015,SinghBeeshamIJGMMP2018,SinghBeeshamIJMPD2019}
(see also our recent work \cite{SinghBeeshamGRG2019} and references therein).

In what follows, we reformulate the model considered by Sahoo {\it et al.} \cite{SahooetalEPJP2014} and follow the above mentioned procedure to study the solutions thoroughly. We first find the constraints for a physically realistic scenario, and then explore the physical and geometrical properties of the model. We further investigate the significance of $f(R,T)$ gravity, and compare the outcomes with the model of GR.

\section{The model in $f(R,T)$ gravity }
\label{sec:2}
Before starting let us make clear that $\rho$ and $p$, respectively, are the effective energy density and pressure in Sect. 2. While the energy-momentum tensor (4), is considered in $f(R,T)$ gravity then $\rho$ and $p$ no longer represent the effective energy density and pressure. As we have mentioned in the introduction that due to the coupling of matter and geometry some extra matter appears in the field equation. We may call it coupled matter. Therefore, in the energy-momentum tensor (4) we replace $\rho$ and $p$ with $\rho_m$ and $p_m$, respectively. The energy-momentum tensor (4) thus represents the primary matter. The notations associated to the coupled matter are defined in Sect. 3.3. In this way, we continue treating  $\rho$ and $p$ as the effective energy density and pressure.

The field equations in $f(R, T)=R+2f(T)$ gravity with the system of units  $8\pi G=1=c$, are obtained as \cite{Harkoetal2011,SinghSinghGRG2014}
\begin{equation}
R_{\mu\nu}-\frac{1}{2}R g_{\mu\nu}=T_{\mu\nu}+2 (T_{\mu\nu}+p_\mathfrak{m} g_{\mu\nu})f'(T)+f(T) g_{\mu\nu},
\end{equation}
where a prime stands for the derivative with respect to $T$.

For $f(T)=\lambda T$, i.e., $f(R,T)=R+2\lambda T$, where $T=\rho_\mathfrak{m}-3p_\mathfrak{m}$, the above field equations reduce to
\begin{equation}
R_{\mu\nu}-\frac{1}{2}R g_{\mu\nu}=(1+2\lambda) T_{\mu\nu}+\lambda(\rho_\mathfrak{m}-p_\mathfrak{m}) g_{\mu\nu},
\end{equation}
which for the metric (1), yield
\begin{eqnarray}
  \left(\frac{\dot A}{A}\right)^2+2\frac{\dot A\dot B}{A B}&=&(1+3\lambda)\rho_\mathfrak{m}-\lambda p_\mathfrak{m},\\
2\frac{\ddot A}{ A}+\left(\frac{\dot A}{A}\right)^2&=&-(1+3\lambda)p_\mathfrak{m}+\lambda \rho_\mathfrak{m},\\
\frac{\ddot A}{A}+\frac{\ddot B}{B}+\frac{\dot A \dot B}{AB}&=&-(1+3\lambda)p_\mathfrak{m}+\lambda \rho_\mathfrak{m}.
\end{eqnarray}
One can see that the terms on the right hand side of field equations ``(13)--(15)" and ``(18)--(20)" in the work done by Sahoo {\it et al.} \cite{SahooetalEPJP2014} contain wrong signs.

\subsection{Model for $m\neq3$}

Using (12), (13) in (16) and (17), for $\lambda\neq-1/2$ and $\lambda\neq-1/4$, we obtain
\begin{eqnarray}
  \rho_\mathfrak{m}&=&\frac{1}{1+2 \lambda}\left[\frac{2 \lambda  (m+3)+3}{(1+4\lambda) m^2 t^2}-\frac{\beta ^2 (k m t)^{-\frac{6}{m}}}{3}\right],\\
  p_\mathfrak{m}&=&\frac{1}{1+2 \lambda}\left[\frac{6 \lambda  (m-1)+2 m-3}{(1+4\lambda) m^2 t^2}-\frac{\beta ^2 (k m t)^{-\frac{6}{m}}}{3}\right].
\end{eqnarray}
These are the correct expressions for the energy density and pressure which are different from those obtained by Sahoo {\it et al.} \cite{SahooetalEPJP2014}.

Any physically realistic cosmological model requires a positive energy density. Moreover, the weak energy condition (WEC)\footnote{$\rho\geq0$, $\rho+p\geq0$} must be satisfied. From (19) we note that if the inequalities $1+2 \lambda <0$ and $\left(2 \lambda  (m+3)+3\right)/(1+2 \lambda)(1+4 \lambda)>0$, i.e.,
 \begin{equation}
\lambda <-\frac{1}{2}\;\;\text{and}\;\; 0<m<-\frac{3(1+2 \lambda)}{2 \lambda },
\end{equation}
hold together, then the energy density can always be positive. However, both these inequalities cannot be satisfied simultaneously, which means that the energy density cannot be positive for all the time. Therefore, the WEC cannot hold throughout the evolution. In contrast, the energy density remains always negative if $1+2 \lambda >0$ and $\left(2 \lambda  (m+3)+3\right)/(1+2 \lambda)(1+4 \lambda)<0$, i.e., if
\begin{equation}
-\frac{1}{2}<\lambda <-\frac{1}{4}\;\;\text{and}\;\; 0<m<-\frac{3(1+2 \lambda)}{2 \lambda },
\end{equation}
or
\begin{equation}
  -\frac{1}{4}<\lambda <0\;\;\text{and}\;\; m>-\frac{3(1+2 \lambda)}{2 \lambda }.
\end{equation}
Consequently, the WEC under the above constraints is violated throughout the evolution.

Notwithstanding, for the set of all values of $\lambda$ and $m$ not obeying the constraints (21)--(23), $\rho_\mathfrak{m}$ can be positive either for a finite period of time $0<t<t_\star$, or from a time $t=t_\star$ to the infinite future, where $t=t_\star$ is a ceratin time at when the energy density transits from positive to negative, or vice-versa. However, it is not possible to find $t=t_\star$ explicitly, due to the complicated expression of the energy density. Therefore, we shall now divide our discussion for all those possible values of $\lambda$ and $m$ for which the energy density can be positive either for $0<t<t_\star$ or for $t>t_\star$. It is to be noted that the transition time $t=t_\star$ can be different in each case.

The models with $m<1$ describe accelerated expansion of the universe which may be an early inflation or the present acceleration. If the model with $m<1$ satisfies the WEC during early times, it means that the model is physically viable only to describe early evolution, and the acceleration must correspond to inflation. Similarly, if the WEC holds at late times, then it must be the present acceleration. On the other hand, the models with $m>1$ will describe decelerating phases whenever the WEC holds good. Let us identify these evolutionary phases in the following cases under the constraints for which the model satisfies the WEC.

\subsection*{case (i) $\lambda<-1/2$ and $m>-3(1+2 \lambda)/2 \lambda $}

Since $m$ can not possess any negative values, and the inequality $m>-3(1+2 \lambda)/2 \lambda $ for $\lambda<-1/2$ is true for all positive values of $m$,  the constraints in this case are equivalent to $\lambda<-1/2$ and $m>0$. We find that if $m<3$, the WEC is obeyed for $0<t<t_\star$, and if $m>3$, it is obeyed for $t>t_\star$. Hence, an acceleration ($m<1$) in this case must be early inflation. On the other hand, the models with $m>1$ can describe the decelerated phase of the universe.

\subsection*{case (ii) $-1/2<\lambda<-1/4$ and $m>-3(1+2 \lambda)/2 \lambda $}

In this case, $m$ can assume any smallest value between $0<m<3$ depending on what value is assigned to $\lambda$. Further, if $m<3$, the WEC holds for $t>t_\star$, while if $m>3$, the WEC holds for $0<t<t_\star$. Hence, an acceleration ($m<1$) in this case means the present acceleration. The models with $m>1$ correspond to a decelerated universe.

\subsection*{case (iii) $-1/4<\lambda<0$ and $m<-3(1+2 \lambda)/2 \lambda $}

The lower bound on $\lambda$ implies that $m$ cannot acquire any positive values less than $3$. The WEC under these constraints hold for $t<t_\star$. Hence, the models in this case can describe only decelerated universe.

\subsection*{case (iv) $\lambda>0$ and $m>-3(1+2 \lambda)/2 \lambda $}

Since $m$ cannot be negative,  the constraints in this case are equivalent to $\lambda>0$ and $m>0$. If $m<3$, the WEC holds for $t>t_\star$, while if $m>3$, it holds for $0<t<t_\star$. Therefore, in this case also an acceleration ($m<1$) means  late time acceleration, and the models with $m>1$ describe decelerated phases.\\

The above discussion well demonstrates that the model is capable of describing the whole cosmological evolution. Now the main objective remains to identify the configuration of matter in various decelerated and accelerated phases. The equation of state (EoS) parameter, $\omega_m=p_m/\rho_\mathfrak{m}$ diverges at $t=t_\star$, so it is not worthwhile using it to depict the behavior of the matter.

A self-interacting scalar field with scalar potential, due to the domination of the potential term over the kinetic term, gives rise to a negative pressure for driving super fast expansion during inflation (for detail see \cite{SinghBeeshamIJGMMP2018} and references therein). On the other hand, when the amplitude of the scalar field is small, it behaves like radiation or dust. In addition, when the field enters into a regime in which the potential energy once again takes over from the kinetic energy, it exerts the same stress as a cosmological constant at late times. However, it happens with a different energy density (in comparison to inflation). Thus, a scalar field can be the most appropriate source for our model. In light of this discussion, we consider a scalar field as the matter source in our further discussion. Since we have already ensured the satisfaction of the WEC, which encapsulates the null energy condition (NEC)\footnote{$\rho+p\geq0$}, the primary matter cannot be of phantom type. We shall also see this fact in the following discussion. The main issue is now to make clear whether the primary matter acts as a perfect fluid or as a quintessential DE.

\subsection{Scalar field model}
\label{sec:5}
The energy density and pressure of a minimally coupled normal ($\epsilon=1$) or phantom ($\epsilon=-1$) scalar field, $\phi$ with self-interacting potential, $V(\phi)$, are given by
\begin{eqnarray}
   \rho_\phi&=&\frac{1}{2}\epsilon\dot\phi^2+V(\phi),\\
   p_\phi&=&\frac{1}{2}\epsilon\dot\phi^2-V(\phi).
\end{eqnarray}
Replacing $\rho_m$ with $\rho_\phi$ and $p_m$ with $p_\phi$ in (19) and (20), the kinetic energy and scalar potential are, respectively, obtained as
\begin{eqnarray}
  \frac{1}{2}\epsilon\dot\phi^2&=&\frac{2 }{3 (1+2 \lambda )}\left[\frac{3}{m t^2}-k^2 (k m t)^{-\frac{6}{m}}\right],\\
V(t)&=&-\frac{2 (m-3)}{(1+4 \lambda ) m^2 t^2}.
\end{eqnarray}
Due to the complicated expression of the kinetic energy it is not possible to find an explicit expression for the scalar field in terms of $t$. Nevertheless, we can determine a physically realistic scenario via requiring the positivity of kinetic energy and scalar potential. For $V>0$, we must have
\begin{eqnarray}
   \lambda &>&-\frac{1}{4}; \;\;\text{if}\;\;0<m<3,\\
   \lambda &<&-\frac{1}{4}; \;\;\text{if}\;\;m>3.
\end{eqnarray}
It is worthwhile mentioning here that a positive scalar potential of the scalar field which obeys the WEC is equivalent to satisfying the dominant energy condition (DEC)\footnote{$\rho\geq|p|$, i.e., $\rho\pm p\geq0$}. Now, on the basis of both the above constraints, we divide our discussion into the following two cases:

\subsection*{case (i) $\lambda<-1/4$ and $m>3$}

In case (ii) of Sect. 3.1, we have found that if $-1/2<\lambda<-1/4$ and $m>3$, the WEC  holds only for a finite period of time $0<t<t_\star$. From (26) we see that the kinetic energy for this restricted period can be positive with quintessential scalar field only. In addition, the strong energy condition (SEC)\footnote{$\rho+3p\geq0$} during $0<t<t_\star$ also holds good, which implies that the scalar field acts as a perfect fluid. Similarly, with the reference to case (i) of Sect. 3.1, the models with $\lambda<-1/2$ and $m>3$ obey the WEC for $t>t_\star$. We further see that the kinetic energy at late  times can be positive with quintessential scalar field only. The SEC also holds good at late  times. Thus, in this case also, the scalar field acts as a perfect fluid.

\subsection*{case (ii) $\lambda>-1/4$ and $m<3$}

In case (iii) of Sect. 3.1, we have found that if $-1/4<\lambda<0$, the WEC can hold for $m>3$ only. Therefore, we exclude the case $-1/4<\lambda<0$, and consider $\lambda>0$ only. Further, with reference to case (iv) of Sect. 3.1, we see that the WEC for $\lambda>0$ and $m<3$ is obeyed for $t>t_\star$, which implies that the model in this case is physically viable to describe  late time evolution only. The kinetic energy at late times can be positive with a quintessential scalar field only. The SEC for $t>t_\star$ is also satisfied. Hence, the scalar field acts as a perfect fluid in this case too.\\

Surprisingly, the models with $m<1$ exhibit accelerating expansion of the universe but,  instead of violating the SEC, the scalar field behaves as a perfect fluid. So a natural question arises: what causes the acceleration in this model? This is not however a contradiction because the scalar field (primary matter) is not the only matter source in this model. In other words, the matter given by the energy-momentum tensor (4) is not the effective matter in this study. The various decelerated and accelerated phases discussed in Sect. 3.1 are driven by the effective matter. Therefore, in accelerating models ($m<1$), the effective matter must violate the SEC. The behavior of the effective matter has already been studied by us elsewhere \cite{SinghBeeshamGRG2019}. In present study, there is another matter source which arises due to the coupling between matter and $f(R,T)$ gravity (cf the right hand side of the field equations (16)--(18)). In the next section, we shall extract the coupling matter from the primary matter and study its nature explicitly. One thing what we can predict about this extra matter at this stage is, when $m<1$ it must mimic the DE which dominates over the primary matter (perfect fluid) and generates the necessary repulsive force to accelerate the universe. In case (iv) of the following section, we shall see that our prediction is true. On the other hand, in the case $m>1$, the coupling matter may behave as DE or a  perfect fluid. This phenomenon can be seen in cases (i)--(iii) of the following section.

\subsection{The behavior of coupled matter}

As we have mentioned above, the matter having energy density $\rho_\mathfrak{m}$ and $p_\mathfrak{m}$ is not the effective matter in this model. Indeed, the coupled terms on the right hand side of equations (16)--(18) containing the parameter $\lambda$ of $f(R,T)$ gravity also contribute as a matter source. Therefore, we separate these extra terms from the energy density and pressure of the matter.

The field equations (16)--(18) can be written as
\begin{eqnarray}
  \left(\frac{\dot A}{A}\right)^2+2\frac{\dot A\dot B}{A B}&=&\rho_\mathfrak{m}+\rho_f,\\
2\frac{\ddot A}{ A}+\left(\frac{\dot A}{A}\right)^2&=&-(p_\mathfrak{m}+p_f),\\
\frac{\ddot A}{A}+\frac{\ddot B}{B}+\frac{\dot A \dot B}{AB}&=&-(p_\mathfrak{m}+p_f),
\end{eqnarray}
where $\rho_f=\lambda(3\rho_\mathfrak{m}-p_\mathfrak{m})$ and $p_f=\lambda(3p_\mathfrak{m}-\rho_\mathfrak{m})$, respectively, are the energy density and pressure of a matter contributed by the coupling between the primary matter and $f(R,T)$ gravity, and are found as
\begin{eqnarray}
  \rho_f&=&\frac{2 \lambda}{3 (2 \lambda +1)}  \left[\frac{3 (12 \lambda -m+6)}{(4 \lambda +1) m^2 t^2}-\beta ^2 (k m t)^{-6/m}\right],\\
  p_f&=&-\frac{2 \lambda }{3 (2 \lambda +1)} \left[\frac{3 (12 \lambda -8 \lambda  m-3 m+6)}{(4 \lambda +1) m^2 t^2}+\beta ^2 (k m t)^{-6/m}\right].
\end{eqnarray}
Let us call these expressions  the energy density and pressure of the coupled matter. The EoS parameter, $\omega_f=p_f/\rho_f$ diverges at $t=t_\star$, so it is not worthwhile using it to depict the behavior of coupled matter. Therefore, we examine the nature of coupled matter through the energy conditions for which we need
\begin{eqnarray}
  \rho_f+p_f&=&\frac{2 \lambda }{3 (2 \lambda +1)} \left[\frac{3 (-12 \lambda +8 \lambda  m+3 m-6)}{(4 \lambda +1) m^2 t^2}-\beta ^2 (k m t)^{-\frac{6}{m}}\right],\\
  \rho_f-p_f&=&-\frac{8 \lambda  (m-3)}{(4 \lambda +1) m^2 t^2}.
\end{eqnarray}
If the coupled matter satisfies the NEC, then from (36) we find that it will satisfy the DEC under the constraints
\begin{eqnarray}
  \lambda &<&-\frac{1}{4} \;\; \text{or}\;\; \lambda >0;\;\; \text{if}\;\; 0<m<3, \\
  -\frac{1}{4}&<&\lambda <0;\;\; \text{if}\;\; m>3.
\end{eqnarray}
The violation of the NEC and the WEC lead to the violation of other energy conditions. Therefore, in case of violation of both of these two, there is no need of examining the SEC and DEC exclusively. On the other hand, if the coupled matter satisfies the WEC then we shall simply call it ``extra matter", or otherwise ``exotic matter", which may be quintessence or phantom. Keeping in view that the WEC must hold for the primary matter, we shall now study the behavior of coupled matter for the restricted times and constraints obtained in cases (i)--(iv) of Sect. 3.1.

\subsection*{case (i) $\lambda<-1/2$ and $m>-3(1+2 \lambda)/2 \lambda $}

The constraints in this case are equivalent to  $\lambda<-1/2$ and $m>0$. If $m<3$, the NEC and WEC for $0<t<t_\star$ are violated. Similarly, if $m>3$, both energy conditions are also violated for $t>t_\star$. Consequently, the coupled matter acts like some exotic matter.

\subsection*{case (ii) $-1/2<\lambda<-1/4$ and $m>-3(1+2 \lambda)/2 \lambda $}

If $m<3$, the NEC and WEC for $t>t_\star$ are violated. Similarly, the NEC and WEC are also violated during $0<t<t_\star$ for $m>3$. Therefore, in this also the coupled matter behaves as exotic matter.

\subsection*{case (iii) $-1/4<\lambda<0$ and $m<-3(1+2 \lambda)/2 \lambda $}

We have seen that the constraints in this case imply $m>3$. The WEC, SEC and DEC for $t>t_\star$ hold good. Hence, the coupled matter contributes as a perfect fluid in this case.

\subsection*{case (iv) $\lambda>0$ and $m>-3(1+2 \lambda)/2 \lambda $}

The constraints in this case are equivalent to $\lambda>0$ and $m>0$. If $m<3$, the WEC is satisfied for $t>t_\star$. As we predicted in case (i) of Sect. 3.2, if $m<1$, the coupled matter must mimic DE in this case, i.e., it must violate the SEC at late times. We see here that the coupled matter violates the SEC not only at late times, but throughout the evolution. Moreover, it also satisfies the DEC. Hence, the coupled matter acts as quintessential DE which dominates over the perfect fluid at late times and provides sufficient repulsion to accelerate the expansion. In other words, it is $f(R,T)$ gravity which is responsible for  late acceleration in this model. On the other hand, although the SEC for $1<m<3$ is also violated, the model decelerates in this case. It is however not a contradiction, again because the primary matter which acts as the perfect fluid must  dominate over the quintessential DE in this case. It is interesting to note that the kinematical behavior of the universe in this case is similar to case (ii) discussed above. However, this case is more realistic physically because the WEC holds for the primary matter, while it is violated in case (ii). For $m>3$, the WEC, SEC and DEC during $0<t<t_\star$ hold good, hence, the coupled matter behaves like a perfect fluid. The universe decelerates for $m>1$ in the presence of effective matter.

\subsection{Model with $m=3$}

The model in this case describes a decelerated universe. The energy density and pressure of primary matter become equal, i.e.,
\begin{eqnarray}
  \rho_\mathfrak{m}=  p_\mathfrak{m}=\frac{ \beta^2- 9k^2}{27 (1+2\lambda)k^2 t^2}.
\end{eqnarray}
For a physically realistic scenario, we must have $\beta^2\geq9k^2$ if $\lambda>-1/2$, and $\beta^2\leq9k^2$ if $\lambda<-1/2$.
The energy density and pressure of the coupled matter also become equal, i.e.,
\begin{eqnarray}
  \rho_f=  p_f=\frac{2 \lambda  \left(9 \beta ^2-k^2\right)}{27 \beta ^2 (1+2 \lambda ) t^2}.
\end{eqnarray}
In view of the constraints for the positivity of primary matter, a realistic cosmological scenario is possible only if $\lambda>0$ when $\beta^2\geq9k^2$, and $\beta^2\leq9k^2$ if $\lambda<-1/2$. It is also to be noted that $\rho_f=2\lambda \rho_\mathfrak{m}$
and $p_f=2\lambda p_\mathfrak{m}$.

Thus, the solutions in this case represent a stiff matter phase of the universe. We note that while the solutions in GR are physically consistent only for $\beta^2\geq9k^2$, in  $f(R,T)$ gravity the model is physically consistent also when $\beta^2\leq9k^2$. This happens due to the parameter $\lambda$ of $f(R,T)=R+2\lambda T$ gravity. The geometrical behaviors in this case also remains the same as in GR \cite{SinghBeeshamGRG2019}. It is to be noted that these solutions were not reported by Sahoo {\it et al.}  \cite{SahooetalEPJP2014}.

\section{Conclusion}

A plane symmetric Bianchi-I model has been studied in $f(R,T)=R+2\lambda T$ gravity. The solutions have been obtained by considering a Hubble parameter $H=k(A^2B)^{-m/3}$, which gives a constant value of the deceleration parameter, $q=m-1$, where $m\geq0$. Consequently, one may either have a decelerating universe ($m>1$) or an accelerating one ($m<1$). The solutions are valid for all values of $\lambda$ except $\lambda\neq-1/2$ and $\lambda\neq-1/4$. The same formulation was considered earlier by Sahoo {\it et al.} \cite{SahooetalEPJP2014}, but due to a wrong sign in the field equations, their solutions are mathematically and physically invalid. Notwithstanding, the appearance of the wrong sign,  the behavior of the geometrical parameters is not altered. We have shown that the geometrical behavior thus remains the same as in GR.

We have not considered the case when $m=0$ as the solutions in this case are similar to an LRS B-I model \cite{SinghBeeshamApSS2019}. However, we have pointed out that Sahoo {\it et al.} \cite{SahooetalEPJP2014} misunderstood the time of origin of the universe in this case. They assumed the origin of the universe at $t=0$, whereas the model has an infinite past. For details, the readers may refer to  \cite{SinghBeeshamApSS2019,SinghBeeshamGRG2019}.

Instead of presenting the results of their own work, Sahoo {\it et al.} \cite{SahooetalEPJP2014} in their concluding section mentioned some general features of $f(R,T)$ gravity. The authors merely added  that the models represent the accelerated expansion of the universe. However, the model with $m>1$ also comprises the decelerated phase as well. Indeed, an acceleration or a deceleration is also not an outcome of their study. Rather, it is guaranteed from the assumption they used to obtain the solutions. In fact, the whole kinematical dynamics of the model in such a formulation becomes independent of $f(R,T)$ gravity. Thus, although the authors considered $f(R,T)$ gravity, they did not discuss the role of modified gravity. Similarly, many others \cite{SharifZubair2013arxiv,ShamirJETP2014,TiwariBeeshamASS2018,ShamirEPJC2015,EsmaeiliJHEP2018} have  considered $f(R,T)=R+2\lambda T$ gravity models, but the role of $f(R,T)$ gravity was not investigated in these works. A common fact in all of these studies is what we noted, viz., the matter violates the NEC. Consequently, no perfect fluid is present in these models. However, the existence of normal matter cannot be neglected.

If one considers any matter in this theory, then due to the coupling between matter and $f(R,T)$ gravity, some extra terms appear on the right hand side of the field equations. These terms must be treated as matter and may be called  coupled matter \cite{HoundjoPiattella2012,Alvarengaetal2013jmp4,Sharifetal2013epjp128,Chakraborty2013,SharifZubair2013jpsj,SinghSinghGRG2014}. It may act either as a perfect fluid or DE. Therefore, the effective matter in these models is a sum of primary matter and  coupled matter. Hence, ensuring that the WEC must hold, the primary matter can be treated as a perfect fluid. We have followed this criteria in the present study. We have done a full treatment to study the properties of the primary matter as well as the coupled matter. Encountering  the dual nature of the primary matter (perfect fluid or exotic matter), we have replaced it with a scalar field (quintessence or phantom) to discriminate between a perfect fluid or exotic matter. Since we have already ensured the NEC, the model becomes consistent only with a quintessence scalar field which may act as a perfect fluid or DE. It is important to mention here that $f(R,T)$ gravity does not alter the behavior of effective matter in the formulations where the kinematical behaviour is fixed by some geometrical parameters. The effective matter in the present study thus also remains the same as in GR \cite{SinghBeeshamGRG2019}. The findings of the present work are summarized as follows:

\begin{description}
  \item[(i) When $\lambda<-1/2$:] The models with $m<1$ can describe early inflation. If $m<3$, the primary matter violates the DEC. Consequently, the model comprises an inflationary epoch at a cost of a violation of the DEC for primary matter. Thus, when we assume that the primary matter is a scalar field, then the corresponding negative scalar potential makes an inflationary scenario physically unrealistic in this model. The coupled matter violates the NEC and WEC for $m<3$ as well as for $m>3$. On the other hand, compliance of the DEC for $m>3$ confirms that the primary matter is a perfect fluid. Therefore, the decelerated models in this case for $m>3$ are physically more viable than those ones with $1<m<3$.

  \item[(ii) When $-1/2<\lambda<-1/4$:] An accelerating expansion ($m<1$) in this case means  late time acceleration. The primary and coupled matter both behave similar to case (i) discussed above. Hence, in this case a late time accelerated model  ($m<1$) is not physically viable. The findings of the decelerated models ($m>1$) are same as mentioned in above point (i).

  \item[(iii) When $-1/4<\lambda<0$:] The models in this case are feasible only for $m>3$. Hence, the model in this case can describe only decelerated universe. Although, the primary matter violates the DEC, the coupled matter satisfies all the energy conditions, and hence contributes as a perfect fluid.  Thus, the models in this case are physically viable.

  \item[(iv) When $\lambda>0$:] An acceleration ($m<1$) in this case means the present accelerating phase. The primary matter is a perfect fluid as it passes all energy conditions. The coupled matter obeys the WEC as well as the DEC for $m<3$, but violates the SEC throughout the evolution. Hence, $f(R,T)$ gravity works as an alternative to quintessential DE which becomes significant at late times and causes an accelerating expansion. On the other hand, when $m>3$ the perfect fluid and coupled matter satisfy all energy conditions. Hence, the coupled matter just works as an additional source of a  perfect fluid. Interestingly, the models in this case, whether decelerated or accelerated, all are physically viable.
\end{description}

\noindent Sahoo {\it et al.} \cite{SahooetalEPJP2014} neglected the solutions for $m=3$. We have obtained the solutions for this case also. The solutions straight forwardly lead to the stiff matter era. Interestingly, while the solutions in GR in this particular case are physically consistent only for $\beta^2\geq9k^2$, $f(R,T)$ gravity makes the model physically consistent also for $\beta^2\leq9k^2$ when $\lambda<-1/2$. The geometrical behaviors in this case also remains the same as in GR \cite{SinghBeeshamGRG2019}.\\

\noindent Since the model is capable of explaining  late time acceleration without the use any hypothetical exotic matter,  $f(R,T)=R+2\lambda T$ gravity can be a good alternative   to GR.

\paragraph{\textbf{Acknowledgement}}
We are thankful to the reviewer for showing his deep interest in this work. His suggestions and constructive comments were very helpful to improve the presentation of the results. Vijay Singh expresses his sincere thank to the University of Zululand, South Africa, for providing a postdoctoral fellowship and necessary facilities. This work is based on the research supported wholly/in part by the National Research Foundation of South Africa (Grant Numbers: 118511).

%

\end{document}